%
%
%
\documentstyle{amsppt}
\newread\epsffilein    
\newif\ifepsffileok    
\newif\ifepsfbbfound   
\newif\ifepsfverbose   
\newdimen\epsfxsize    
\newdimen\epsfysize    
\newdimen\epsftsize    
\newdimen\epsfrsize    
\newdimen\epsftmp      
\newdimen\pspoints     
\pspoints=1bp          
\epsfxsize=0pt         
\epsfysize=0pt         
\def\epsfbox#1{\global\def\epsfllx{72}\global\def\epsflly{72}%
   \global\def\epsfurx{540}\global\def\epsfury{720}%
   \def\lbracket{[}\def\testit{#1}\ifx\testit\lbracket
   \let\next=\epsfgetlitbb\else\let\next=\epsfnormal\fi\next{#1}}%
\def\epsfgetlitbb#1#2 #3 #4 #5]#6{\epsfgrab #2 #3 #4 #5 .\\%
   \epsfsetgraph{#6}}%
\def\epsfnormal#1{\epsfgetbb{#1}\epsfsetgraph{#1}}%
\def\epsfgetbb#1{%
%
%
\openin\epsffilein=#1
\ifeof\epsffilein\errmessage{I couldn't open #1, will ignore it}\else
%
%
   {\epsffileoktrue \chardef\other=12
    \def\do##1{\catcode`##1=\other}\dospecials \catcode`\ =10
    \loop
       \read\epsffilein to \epsffileline
       \ifeof\epsffilein\epsffileokfalse\else
%
%
          \expandafter\epsfaux\epsffileline:. \\%
       \fi
   \ifepsffileok\repeat
   \ifepsfbbfound\else
    \ifepsfverbose\message{No bounding box comment in #1; using defaults}\fi\fi
   }\closein\epsffilein\fi}%
%
%
\def\epsfclipstring{}
\def\epsfsetgraph#1{%
   \epsfrsize=\epsfury\pspoints
   \advance\epsfrsize by-\epsflly\pspoints
   \epsftsize=\epsfurx\pspoints
   \advance\epsftsize by-\epsfllx\pspoints
%
%
   \epsfxsize\epsfsize\epsftsize\epsfrsize
   \ifnum\epsfxsize=0 \ifnum\epsfysize=0
      \epsfxsize=\epsftsize \epsfysize=\epsfrsize
      \epsfrsize=0pt
%
%
     \else\epsftmp=\epsftsize \divide\epsftmp\epsfrsize
       \epsfxsize=\epsfysize \multiply\epsfxsize\epsftmp
       \multiply\epsftmp\epsfrsize \advance\epsftsize-\epsftmp
       \epsftmp=\epsfysize
       \loop \advance\epsftsize\epsftsize \divide\epsftmp 2
       \ifnum\epsftmp>0
          \ifnum\epsftsize<\epsfrsize\else
             \advance\epsftsize-\epsfrsize \advance\epsfxsize\epsftmp \fi
       \repeat
       \epsfrsize=0pt
     \fi
   \else \ifnum\epsfysize=0
     \epsftmp=\epsfrsize \divide\epsftmp\epsftsize
     \epsfysize=\epsfxsize \multiply\epsfysize\epsftmp   
     \multiply\epsftmp\epsftsize \advance\epsfrsize-\epsftmp
     \epsftmp=\epsfxsize
     \loop \advance\epsfrsize\epsfrsize \divide\epsftmp 2
     \ifnum\epsftmp>0
        \ifnum\epsfrsize<\epsftsize\else
           \advance\epsfrsize-\epsftsize \advance\epsfysize\epsftmp \fi
     \repeat
     \epsfrsize=0pt
    \else
     \epsfrsize=\epsfysize
    \fi
   \fi
%
%
   \ifepsfverbose\message{#1: width=\the\epsfxsize, height=\the\epsfysize}\fi
   \epsftmp=10\epsfxsize \divide\epsftmp\pspoints
   \vbox to\epsfysize{\vfil\hbox to\epsfxsize{%
      \ifnum\epsfrsize=0\relax
        \includegraphics{#1}%
      \else
        \epsfrsize=10\epsfysize \divide\epsfrsize\pspoints
        \includegraphics{#1}%
      \fi
      \hfil}}%
\global\epsfxsize=0pt\global\epsfysize=0pt}%
%
%
{\catcode`\%=12 \global\let\epsfpercent=
%
%
\long\def\epsfaux#1#2:#3\\{\ifx#1\epsfpercent
   \def\testit{#2}\ifx\testit\epsfbblit
      \epsfgrab #3 . . . \\%
      \epsffileokfalse
      \global\epsfbbfoundtrue
   \fi\else\ifx#1\par\else\epsffileokfalse\fi\fi}%
%
%
\def\epsfempty{}%
\def\epsfgrab #1 #2 #3 #4 #5\\{%
\global\def\epsfllx{#1}\ifx\epsfllx\epsfempty
      \epsfgrab #2 #3 #4 #5 .\\\else
   \global\def\epsflly{#2}%
   \global\def\epsfurx{#3}\global\def\epsfury{#4}\fi}%
%
%
\def\epsfsize#1#2{\epsfxsize}
%
%

%
%
\nologo
\hsize 32pc
\vsize 50pc
\emergencystretch=100pt

\def\ms{{\medskip}}
\def\del{{\partial}}
\def\k{{\kappa}}
\def\t{{\tau}}

\def\slf{{\Cal F}}
\def\slg{{\Cal G}}

\def\slr{{\Cal R}}

\def\slf{{\Cal F}}

\def\slp{{\Cal P}}

\def\smalltype{\let\rm=\eightrm \let\bf=\eightbf
\let\it=\eightit \let\sl=\eightsl \let\mus=\eightmus
\baselineskip=9.5pt minus .75pt \rm}
\parindent=30pt

\def\onepsi{\hbox{\hbox{$^1$}\kern-.15em $\Psi$}}
\def\onejt{\hbox{\hbox{$^1$}\kern-.25em $\tilde J$}}
\def\onext{\hbox{\hbox{$^1$}\kern-.25em $\tilde X$}}
\def\oneit{\hbox{\hbox{$^1$}\kern-.25em $\tilde I$}}
\def\onert{\hbox{\hbox{$^1$}\kern-.1em $\tilde \slr$}}
\def\oneslr{\hbox{\hbox{$^1$}\kern-.1em $\slr$}}

\def\udots{\mathinner{\mkern1mu\raise1pt\vbox{\kern7pt\hbox{.}}\mkern2mu
\raise4pt\hbox{.}\mkern2mu\raise7pt\hbox{.}\mkern1mu}}
\def\uudots{\mathinner{\mkern1mu\raise2pt\vbox{\kern7pt\hbox{.}}\mkern2mu
\raise5pt\hbox{.}\mkern2mu\raise8pt\hbox{.}\mkern1mu}}

\magnification =1200
\def\myprime{^\prime}
\pagewidth{ 12 cm }
\pageheight{ 7.5 in}
\baselineskip=16pt plus 2pt
\font\twelverm=cmr12
%
%
\topmatter

\title
{Localized Induction Hierarchy} \\
{and}  \\
{Weingarten systems}
\endtitle
\author
Ron Perline 
\endauthor
\affil
Dept. of Mathematics and Computer Science, Drexel University
\endaffil
\abstract
We describe a method of constructing Weingarten
systems of triply orthogonal coordinates, related
to the localized induction equation hierarchy of
integrable geometric evolution equations.
$$ \quad $$
{\twelverm
\centerline{ Submitted to } 
\centerline{}
\centerline{ Physics Letters A}
}
$$\quad $$
\endabstract
\endtopmatter

\leftheadtext{LIH and Weingarten systems}

\newpage

{\bf 1. Introduction}.
\medskip
Recently, there has been a great deal of research related
to problems from classical differential geometry, approached
from the modern perspective of integrable systems (or soliton
theory).  As representative works we mention those of Sym, Pinkall, 
Bobenko, and Melko and Sterling
[Sym], [Bur], [Bob], [Mel-S].

In a   related series of papers
[Lan-P 1-5], [Per],
we have studied
the {\it localized induction equation} (LIE)  and have
investigated its role in differential geometry.  LIE is
a local geometric evolution equation defined on space curves
via the equation
$$ \gamma_t = \gamma_s \times \gamma_{ss}$$
where $s$ is arclength parameter for an evolving space
curve $\gamma(s,t) \in R^3$.  LIE can also be written 
as $\gamma_t = \kappa B$, where $\kappa$ is the curvature
and $B$ is the binormal of the curve.  LIE is an extremely
idealized model for the evolution of the centerline of a thin,
isolated vortex tube in an inviscid fluid; it was first developed
at the turn of the century by da Rios, a student of the
geometer  Levi-Civita,
and rediscovered some thirty years ago (for derivation and history,
see [Bat],[Lam 1],[Ric 1]).Thus the roots of LIE
are historically grounded in the
boundary between mathematical physics  and geometry.

Hasimoto [Has]  found  the connection between LIE and soliton theory:
if $\gamma$ evolves according to LIE, then Hasimoto showed that the
associated {\it complex curvature function}
$\psi(s)  = \kappa(s) e^{i \int^s \tau(u) \, du}$ ($\tau$ is the curve's
torsion)
evolves according to the {\it cubic nonlinear Schr\"odinger
equation} (NLS) 
$\psi_t = i(\psi_{ss} + {1 \over 2}|\psi|^2 \psi)$.  NLS is
one of the standard examples of an integrable partial differential
equation; the result of Hasimoto states that LIE is a geometric
realization of NLS. We  remark that there is now machinery for
producing geometric realizations of integrable equations via
the {\it soliton surface} approach [Sym].  In particular, This method
indeed ``produces" LIE as a geometric realization of NLS.

As a consequence of the LIE-NLS corresondence, LIE inherits the
characteristics associated with integrable equations:  soliton
solutions, an infinite sequence of conserved Hamiltonians in
involution, and  a  corresponding hierarchy of commuting Hamiltonian 
vector fields.
We refer to  this hierarchy as the {\it localized induction hierarchy} (LIH).

As a curve evolves through space with a velocity field which is a member of LIH,
it sweeps out a surface; and one can ask if there is any interesting
structure to the resulting surface.  For example,
if the curve $\gamma(s,t)$ evolves according to LIE, then it is
easy to see that $\gamma(s, t_0)$ is a geodesic on the resulting
surface, for any fixed time $t_0$. 

In a similar fashion, we have recently constructed {\it pseudospherical
surfaces} (= surface of constant negative  Gauss curvature)
using the integrability properties of LIH [Per].  We demonstrated
that certain {\it soliton curves} (= critical points for linear combinations
of the conserved Hamiltonians), evolving according to linear combinations
of the vector fields from LIH, sweep out pseudospherical surfaces.

In this paper, we extend this result to give a dynamical prescription 
for producing {\it Weingarten systems}.  A Weingarten system is a triply
orthogonal system of surfaces, such that the surfaces in one family
are pseudospherical [Eis].  The subject of triply orthogonal systems
is an old one in mathematical physics and geometry, which continues
to have interesting developments [Mil]. The connection we make between
Weingarten systems and integrable geometric evolution equations 
seems to be new.

We begin by  reviewing relevant facts about the structure
of LIH and related vector field hierarchies  needed for our construction.
We then describe the procedure for constructing Weingarten systems.
To keep the paper self-contained,
we briefly include some material from an earlier
paper.

\newpage

{\bf 2. LIH and related hierarchies}.
\medskip
As stated in the introduction, LIE belongs to an infinite hierarchy of 
commuting evolution
equations on curves, all of the form $\gamma_t = X_n = aT + bN + cB$, where
$\{ T,N,B \}$ is the Frenet frame along the curve,
and $a,b,c$ are functions (polynomial) of   $\kappa, \tau, \kappa' =
\kappa_s, \tau' = \tau_s$, and higher derivatives with respect to $s$.
We list the first 
few terms of the hierarchy, as well as their associated
Hamiltonians:
$$
\eqalign{
& X_{1} = \kappa B, \quad  I_{1} = \int_\gamma \ ds , \cr
& X_{2} = {{\kappa^2} \over 2 }T  +   \kappa \myprime N + \kappa \tau B, 
 \quad I_{2} = \int_\gamma -\tau \ ds ,\cr
& X_{3} \ = 
\kappa^2 \tau T 
+ (2 \kappa\myprime \tau + \kappa \tau \myprime)N 
+ (\kappa \tau^2 - \kappa ^{\prime \prime} -{1 \over 2}\kappa^3)B ,  \cr
& \quad I_3 = \int_\gamma {1 \over 2} \kappa^2 \ ds , \cr
& X_{4} \ =
(-\k \k '' + {1 \over 2} (\k ')^2 + {3 \over 2} \k ^2 \t ^2 - {3 \over 8} \k
^4)T \cr
& + ( - \k ''' + 3 \k \t \t ' + 3 \k ' \t ^2 - {3 \over 2} \k ^2 \k ') N \cr
& + ( \k \t ^3 - 3 (\k ' \t ) ' -  {3 \over 2} \k ^3 \t - \k \t '') B, \cr
& \quad I_{4} = \int_\gamma  \, {1 \over 2} \kappa^2 \tau \,{ds}, \cr
& \dots \cr
}
$$
The vector fields of LIH are {\it locally arclength preserving}
: a vector field $W$ is locally archlength preserving 
if every segment of a curve $\gamma$
has its length remain constant as $\gamma$ evolves via $\gamma_t = W$.
Equivalently, $<W_s,T> = 0$.  A discussion of the physical interpretation
of some of these vector fields and associated functionals can be found
in [Ric 2].

LIH is generated
by a recursion operator $X_{n+1} = \slr X_n, \ n \ge 0$; if
$X = aT + bN + cB$ then $\slr(X) = -\slp(T \times X')$, where 
$\slp$ is a {\it parameterization operator}
$\slp(X) = {\int^s (\kappa b) ds}\, T + bN + cB$.
Using $\slr$,  we can  compactly express
the first-order variations in curvature and 
torsion along {\it any} vector field
$W$ which is locally arclength preserving [Lan-P 3]: 
$$W(\kappa) = <-\slr^2(W),N>,$$
$$W(\tau) = < -\slr^2(W),B/\kappa >' \, .$$

There are a number of hierarchies of integrable geometric evolution
equations, related to LIE,  which have interesting geometric properties.
These are discussed in more detail in [Lan-P 3].  We recall two which
are relevant to our construction of Weingarten systems:
\newline
(1) {\it Constant torsion preserving}:  For $n \ge 0$, the
vector fields 
$$Z_n = \sum_{k=0}^{2n-1} {{\binom {2n+1}k} (-\tau_0)^k X_{2n-k}}$$
preserve the constant torsion condition $\tau = \tau_0$.
If a constant torsion curve $\gamma$ evolves according to
$\gamma_t = Z_n$, the induced evolution on curvature
$\kappa_t = Z_n(\kappa)$ is the corresponding element
of the (mKdV) hierarchy; in particular, $Z_1$ induces the
(mKdV) evolution $\kappa_t = \kappa_{sss} + {3 \over 2} \kappa^2 \kappa_s$,
recovering a result of Lamb [Lam 2].
\newline
(2) {\it Torsion independent}: The 
vector fields 
$$A_n = \sum_{k=0}^{n-1} {{\binom {n-1}k} (-\tau_0)^k X_{n-k}} ,
\quad n \ge 1$$
have the property that, along curves $\gamma$ with $\tau = \tau_0$, the
coefficients of $A_n = aT + bN + cB$ have no explicit $\tau$ dependence.
The odd vector fields in the sequence are purely binormal; the even
vector fields, on the other hand, have zero binormal component. We thus
refer to the  odd fields as ``binormal"
and the even fields as ``planar-like" (since for planar curves they  are
tangent to the ambient plane).
\newline
\newline
{\bf 3.  Planar-like soliton  curves and their Killing fields}
\medskip
We define a {\it planar-like $n$-soliton curve} to be a curve $\gamma$
such that $\gamma$ has constant torsion $\tau = \tau_0$ and 
the vectorfield $\sum_{i=0}^n a_i A_{2i+1}$ vanishes along $\gamma$, for
some choice of $a_0, a_1, \dots, a_n$.  By scaling the $a_i$'s, we
can assume without loss of generality that $a_n =1$. 
The condition $\sum_{i=0}^n a_i A_{2i+1} = 0$  is a nonlinear differential
equation for the curvature $\kappa(s)$ whose coefficients are independent
of $\tau$; thus, planar-like soliton curves have the same curvature function
as some planar curve, and their torsion differs by a constant $\tau_0$ 
(hence the terminology ``planar-like").
It is quite easy to show that the condition  $\sum_{i=0}^n a_i A_{2i+1} = 0$
is equivalent to $\gamma$ being a critical point for an appropriate linear
combination of the conserved Hamiltonians $I_i$.

Previously [Lan-P 2], 
we  have considered the {\it reconstruction problem} for certain
soliton curves:  given the curvature and torsion function for such a curve,
is there an efficient way of integrating the associated Frenet equations
for the curve?  It turns out that a key ingredient is that one can
express certain  {\it Killing
fields} in terms of the local geometric invariants of the soliton curve.
A Killing field is an infinitesimal isometry; that is, it generates
a one-parameter flow of rigid motions.  A vector field $V$ is called 
{\it Killing along $\gamma$} if it extends to a Killing field in $R^3$
(modulo translation along the curve).
Although we do not reconsider the reconstruction problem here, it will
be useful to have formulas for the Killing fields of our planar-like soliton
curves:
\medskip
{\it Proposition:  let $\gamma$ be a planar-like soliton curves as defined
above. Then \newline
(i)  The vector field $ K_1 = \sum_{i=1}^n a_i Z_i$ is Killing along $\gamma$. \newline
(ii)  The condition  $\sum_{i=0}^n a_i A_{2i+1} = 0$ can be rewritten 
$\sum_{j=1}^{2n+1} b_j X_j = 0$, where 
$b_j = \sum_{i = [{{j-1} \over {2}}]}^n a_i {\binom{2i} {j-1}} (-\tau_0)^{2i + 1 - j}$.
The vector field $K_2 = \sum_{j=3}^{2n+1} b_j X_{j-2}$ is Killing along $\gamma$.
}

Using the definition of a planar-like soliton curve, and the variation formulas for
$\kappa$ and $\tau$ given above, it is easy to see that the first order variations
of $\tau$ and $\kappa$ in the direction of the fields $K_1, K_2$ are $0$ and $\alpha \, \kappa'(s)$
respectively, where $\alpha$ is some constant.  That the variations in $\tau$ and $\kappa$ be
of this form is obviously  a {\it necessary} condition for the fields $K_1, K_2$ to be Killing
along
$\gamma$; an argument of Langer and Singer [Lan-S] 
shows that it is in fact sufficient.
\newline
\newline
{\bf 4.  Vector fields associated to a planar-like soliton curve}.
\newline
We now state the technical proposition which allows us to construct
Weingarten systems:
\medskip
{\it Proposition: Let $\gamma$ be a planar-like soliton curve. Then: \newline
(i) The planar-like vector field $\slf = T(x) + (-1/a_0)(\sum_{i=0}^n a_i A_{2i})$ 
is constant torsion preserving along $\gamma$.  $\slf$ is a unit
vector field which asymptotically ($s \rightarrow {\pm \infty}$)
equals $T(x)$. \newline
(ii) The binormal vector field $\slg = \sum_{i=0}^{n-1} e_i A_{2i+1}$ is 
constant
torsion preserving along $\gamma$, where
$e_i = \sum_{k=0}^{n-i-1} a_{k+i+1} (-1)^k {\tau_0}^{2k}$.
}
\newline
The vector field $\slf$ is the same as the vector field $T^{*}$ introduced
in [Per].  As stated there, to show $\slf$ is constant torsion preserving,
it suffices to prove the identity
$$  \slf =  T(x) + {{-1} \over {a_0 \tau_0^2}}{\sum_{k=0}^{n+1} (a_{k-1} -2a_k\tau_0^2 + a_{k+1}\tau_0^4)Z_k} ,$$
along $\gamma$, thus expressing $\slf$ as a linear combination of vector fields known to preserve the
constant torsion condition.

Similarly, to prove that $\slg$ is constant torsion preserving, one simply verifies
the identity
$\slg =  K_2 + {\sum_{j=1}^{n-1} g_j Z_j}$, where
$$g_j = 2 \sum_{i=0}^{n-j-1} a_{n-i} (-1)^{i+j} \tau_0^{2n -2i -2j -1} \ . $$
The vector fields  $Z_j$ are constant torsion preserving along any curve with
$\tau = \tau_0$;
and since $K_2$ is Killing, it also is constant torsion preserving
along $\gamma$.
\newline
\newline
{\bf 5.  Weingarten systems}.
We are now in a position to construct Weingarten systems in a simple manner.
Start with a planar-like $n$-soliton curve $\gamma_0(s)$.
Consider the evolution equations
$\gamma_{t_1} = \slf$, $\gamma_{t_2} = \slg$,
both with initial conditions $\gamma(0,s) = \gamma_0(s)$.
The resulting flows commute, and both are locally arc-length preserving.
Thus it makes
sense to discuss $\gamma(s,t_1, t_2)$ which generically 
(away from self-intersections)  defines
a coordinate system for some subset $U$ of $R^3$;  
here,  $\gamma(s,t_1, t_2)$  denotes the image of the initial
curve  $\gamma_0$ 
after evolving $t_1$ time units via the first evolution equation,
and $t_2$ time units via the second. 

As discussed above, $\gamma_0$ is a critical curve  for an 
appropriate linear combination
of the conserved functionals for LIH.
The set of such  critical curves is conserved by
the evolutions associated with the vector fields $\slf,
 \slg$, so for any $t_1, t_2$
the curve $\gamma(s, t_1, t_2)$
is a critical curve.  Similarly, the constant torsion condition $\tau = \tau_0$
is preserved, so  $\gamma(s, t_1, t_2)$ is an planar-like soliton curve 
for all ``times" $t_1, t_2$.

At a generic point in the set $U$, we have the three tangent vectors
${{\del} \over {\del s}} = T, {{\del} \over {\del t_1}} = \slf, {{\del} \over {\del t_2}} =
\slg$.  By construction, $\slg$, which is binormal, is perpendicular to the planar-like field
$\slf$ and the unit tangent field $T$.  

For fixed $t_2$, we can refer to the results of [Per] to 
conclude that the level surface
$\gamma(s,t_1, t_2)$ is pseudospherical; in fact, $s, t_1$ are just the 
{\it asymptotic coordinates}
for the pseudospherical surface.  Finally, we can construct orthogonal coordinates for the
pseudospherical surface by using the {\it principal curvature coordinates} 
$r_1  = (s + t_1)/2, \, r_2  = (s  - t_1)/2$.  
It is clear that the coordinate system
$\{r_1, r_2, t_2\}$ is a Weingarten system.

As an example of our construction, we give explicitly
the associated one$-$soliton Weingarten system.  We start
with the well-known Hasimoto filament [Has]:
$$
x = s-{\frac {2\,\nu\,\tanh(\nu\,s)}{\nu^{2}+\tau_0^{2}}},
y= {\frac {2\,\nu\,\hbox{sech}(\nu\,s)
\cos(\tau_0\,s)}{\nu^{2}+\tau_0^{2}}},$$
$$z = {\frac {2\,\nu\,\hbox{sech}(\nu\,s)\sin(\tau_0\,s)}
{\nu^{2}+\tau_0^{2}}} \ .$$
This curve has constant torsion $\tau_0$ and curvature
$\kappa(s) =  2 \nu \, \hbox{sech}( \nu s)$, and is critical for
the functional $I_3 -2\tau_0 I_2 + (\tau_0^2 + \nu^2) I_1$.  
For this curve, the vector fields $\slf$ and $\slg$ restrict
to {Killing fields}, which facilitates the integration
of the associated flows.  The resulting Weingarten system
is given by:
$$x = 
{\frac {\nu^{2}\cosh({ s_1}){ r_1}
+\cosh({ s_1})\tau_0^{2}{ r_1}
-2\,\nu\,\sinh({ s_1})}
{\left (\nu^{2}+\tau_0^{2}\right )\cosh({ s_1
})}}, \ 
y = {\frac {2\,\nu\,\left (\cos({ s_2})
\right )}{\left (\nu^{2}+\tau_0^{2}
\right )\cosh({ s_1})}},$$
$$z = {\frac {2\,\nu\,\left (
\sin({s_2})\right )}{\left (\nu^{2}+\tau_0^{2}
\right )\cosh({ s_1})
}},$$
where
$$s_1 =
{\frac {\left ({\frac {{ r_1}}{2}}
+{\frac {{ r_2}}{2}}\right )\nu^
{2}-2\,\tau_0\,t_2\nu^{2}+\left ({\frac 
{{ r_1}}{2}}-{\frac {{ r_2}}{2
}}\right )\tau_0^{2}}{\nu}}, \ 
s_2 =
\nu^{2}t_2 +\tau_0\,r_2-t_2\tau_0^{2}  \ ;$$
the image of the Weingarten system is the interior of a cylinder
in $R^3$.

By our construction, the surfaces corresponding to $t_2 = \ constant$ are 
pseudospherical, and this can be verified directly from the formulas
for the Weingarten system; their  curvature is $-\tau_0^2$.
The other families of surfaces for the
Weingarten system also have interesting properties:  the surfaces
$r_1 = \ constant$ are $spheres$, with curvature 
${\frac {1} {4}}{\frac {(\nu^2 + \tau_0^2)^2} {\nu^2}}$, and the
surfaces $r_2 = \ constant$ are again pseudospherical, with curvature
$-{\frac {1} {4}}{\frac  {(\nu^2 - \tau_0^2)^2} {\nu^2}}$  Our
illustration on the next page  shows intersecting
pieces of three orthogonal surfaces from our families,  corresponding
to the parameter values $\nu = 2, \, \tau_0 = 1$.
\medskip
\newpage
\centerline{
\epsfbox{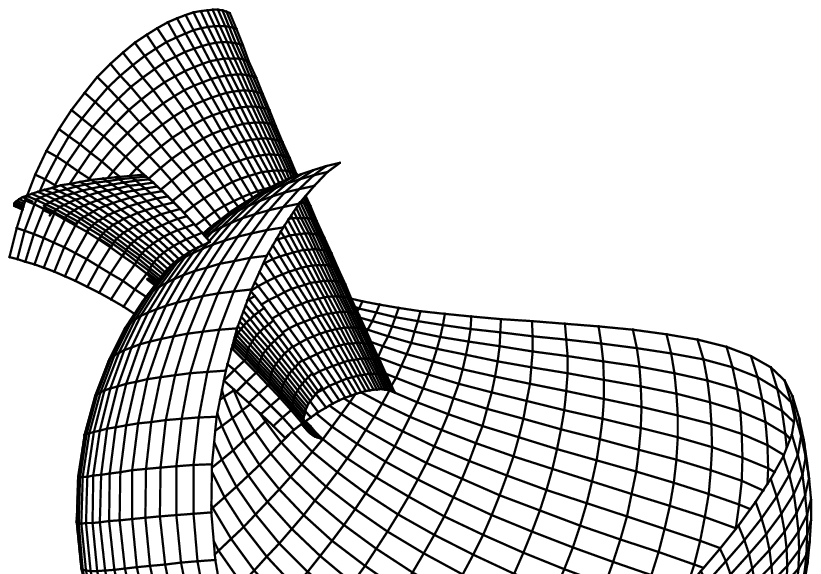}
}
\newpage

{\bf Acknowledgements}.  The connection of Weingarten systems to
integrable systems was brought to our attention by Colin Rogers, whom
we thank for several informative and entertaining conversations.  
Independently, Rogers is developing an alternative approach to
the integrability theory of Weingarten systems (private communication),
which is likely to be more computational effective than the method
described here.
The catalyst for these discussions was the recent ``Bianchi Days"
Workshop of the Warsaw University Physics Department;   we extend
a warm thanks to Antoni Sym and his colleagues for the interesting
and stimulating program they organized.
\newline
\newline
\newline

\end